\documentclass[conference]{IEEEtran}
\IEEEoverridecommandlockouts
\usepackage{cite}
\usepackage{amsmath,amssymb,amsfonts}
\usepackage{algorithmic}
\usepackage{graphicx}
\usepackage{textcomp}
\usepackage{gensymb}
\usepackage{xcolor}
\usepackage{color,soul}
\usepackage{comment}
\usepackage{wrapfig}

\usepackage[T1]{fontenc}
\usepackage[utf8]{inputenc}
\usepackage{authblk}

\ifCLASSOPTIONcompsoc
    \usepackage[caption=false, font=normalsize, labelfont=sf, textfont=sf]{subfig}
\else
\usepackage[caption=false, font=footnotesize]{subfig}
\fi

\usepackage{geometry}

\def\BibTeX{{\rm B\kern-.05em{\sc i\kern-.025em b}\kern-.08em
    T\kern-.1667em\lower.7ex\hbox{E}\kern-.125emX}}

\begin{document}

\title{RESTRAIN: Reinforcement Learning-Based Secure Framework for Trigger-Action IoT Environment}
\author[1,$\dagger$]{Md Morshed Alam \thanks{* Corresponding author, $\dagger$ Equal Contribution}}
\author[2,*,$\dagger$]{Lokesh Chandra Das}
\author[3]{Sandip Roy}
\author[3]{Sachin Shetty}
\author[4]{Weichao Wang}
\affil[1]{School of Cybersecurity, Old Dominion University, Norfolk, US}
\affil[2]{School of Computing, Wichita State University, Wichita, US}
\affil[3]{Center for Secure \& Intelligent Critical Systems, Old Dominion University, Norfolk, USA}
\affil[4]{Department of Software and Information Systems, University of North Carolina at Charlotte, Charlotte, USA}
\affil[ ]{\{m2alam, sroy, sshetty\}@odu.edu, lokesh.das@wichita.edu, wwang22@charlotte.edu}
\maketitle

\begin{abstract}
Internet of Things (IoT) platforms with trigger-action capability allow event conditions to trigger actions in IoT devices autonomously by creating a chain of interactions. Adversaries exploit this chain of interactions to maliciously inject fake event conditions into IoT hubs, triggering unauthorized actions on target IoT devices to implement remote injection attacks. Existing defense mechanisms focus mainly on the verification of event transactions using physical event fingerprints to enforce the security policies to block unsafe event transactions. These approaches are designed to provide offline defense against injection attacks. The state-of-the-art online defense mechanisms offer real-time defense, but extensive reliability on the inference of attack impacts on the IoT network limits the generalization capability of these approaches. In this paper, we propose a platform-independent multi-agent online defense system, namely RESTRAIN, to counter remote injection attacks at runtime. RESTRAIN allows the defense agent to profile attack actions at runtime and leverages reinforcement learning to optimize a defense policy that complies with the security requirements of the IoT network. The experimental results show that the defense agent effectively takes real-time defense actions against complex and dynamic remote injection attacks and maximizes the security gain with minimal computational overhead.

\end{abstract}

\begin{IEEEkeywords}
Internet of Things, Trigger-action Platform, Remote Injection Attack, Reinforcement Learning, Deep Recurrent Q-Network, Multi-Agent System.
\end{IEEEkeywords}

\section{Introduction}
Trigger action capabilities in Internet of Things (IoT) platforms make it easier for users to set up customized rules and let IoT hubs enforce these rules to automate network tasks. For example, a homeowner could set up a rule that instructs an IoT hub to ask a smart sprinkler system to open the water valve when a thermostat measure is $130$\textdegree F considering there is a fire hazard. Users can create functional dependencies and leverage causal relationships among IoT devices to set up rules in a \emph{rule engine} \cite{fan2021_ruleedger}. Smart hubs implement these rules in real-time and invoke actions in IoT devices based on the occurrence of prerequisite event triggers defined in the rules \cite{celik2019a_iotsec_survey}. Note that a smart hub is the central entity in an IoT network through which IoT devices communicate with each other. Typically, IoT devices report their cyber state to the hubs notifying recent event transactions. The hubs verify the physical states of the reporting devices and invoke actions in corresponding IoT devices based on predefined rules \cite{fan2021_ruleedger}. 


An event transaction contains the details of an \emph{event condition} which acts as a \emph{trigger} for the relevant actions in other devices. The execution of user-defined rules creates a \emph{chain of interactions} incorporating a sequence of triggers and corresponding actions \cite{alam2022_iotmonitor}. The hubs utilize the chain of interactions to automate network tasks and report device changes to users. However, an attacker can exploit the chain of interactions to maliciously inject fake event conditions into the hubs to carry out \emph{remote injection attacks} \cite{alam2024iotwarden}. The objective of the attacker is to instantiate invalid actions in IoT devices and compromise the functionality of a goal node (see Section \ref{attack_characterization}) \cite{alam2021_survey}. This type of attack is also referred to as \emph{event spoofing attack} since the attacker injects spoofed event conditions into the hub to implement the attack \cite{ozmen2023-evation-attacks}.


Existing research mainly focuses on offline anomaly detection systems (i.e., event verification systems) to mitigate IoT environment vulnerabilities \cite{ozmen2023-evation-attacks}, \cite{sbirnbach2019_peeves}, \cite{fu2021-hawatcher}, \cite{Celik2019IoTGuardDE}, \cite{nguyen2018_IoTSan}. These systems first verify the occurrence of event conditions based on \emph{physical event fingerprints} \cite{sbirnbach2019_peeves} captured by sensors and utilize rule-based or machine learning based approaches to ensure that the event transactions comply with predefined security policies, thereby blocking the unsafe transactions. These security systems do not offer real-time protection against active remote injection attacks. Additionally, many of them require source code modification of IoT applications that limits the generalization capability of defense solutions \cite{Celik2019IoTGuardDE}. To address the limitations of offline defense mechanisms, some researchers propose single-agent based online security solutions to provide real-time defense against agile remote injection attacks \cite{alam2024iotwarden}, \cite{alam2024-iothaven}. These approaches rely on the impact of malicious activities in the network to infer the attack sequences, thereby applying relevant defense actions. However, the impact of malicious activities may not accurately reflect how attackers intrude the IoT environment. Therefore, the defense policy may not be adequate to effectively counter dynamic and complex attack activities in real-time.

In this paper, we introduce RESTRAIN, a reinforcement learning (RL)-based online security framework that simulates a realistic attack-defense scenario and trains a smart defense agent to guard the IoT environment against dynamically changing remote injection attacks. In RESTRAIN, a defense agent and an attack agent are trained independently to optimize a defense policy and an attack policy, respectively.


We make the following contributions in this paper:
\begin{itemize}
    \item We propose an RL-based multi-agent defense system, called RESTRAIN, that allows a defense agent to model attack activities at runtime and take optimal defense actions to secure IoT networks against progressing remote injection attacks.
    
    \item We design novel reward functions for both attack and defense agents, enabling them to achieve maximized attack and security gains through the optimal selection of attack and defense actions.
    
    \item We implement RESTRAIN using TensorFlow \cite{abadi2016tensorflow} and train attack and defense agents using Deep Recurrent Q-Network (DRQN) \cite{hausknecht2015deep}.  
    
    \item We conduct an extensive experiment to evaluate the performance of RESTRAIN. Simulation results exhibit that RESTRAIN effectively defends against agile remote injection attacks with minimal computational overhead.
\end{itemize}

The rest of the paper is organized as follows. In section \ref{attack_overview}, we provide an overview of a remote injection attack and introduce a relevant threat model. In section \ref{restrain-system-overview}, we describe the proposed defense system in detail. The experimental details and the simulation results are presented in section \ref{simulation-results}. Finally, in section \ref{conclusion}, we conclude the paper and highlight some future research directions.

\section{Attack Overview}
\label{attack_overview}

\begin{figure}[htbp]
    \centering
    \vspace{-5pt}
    \includegraphics[width=0.95\linewidth]{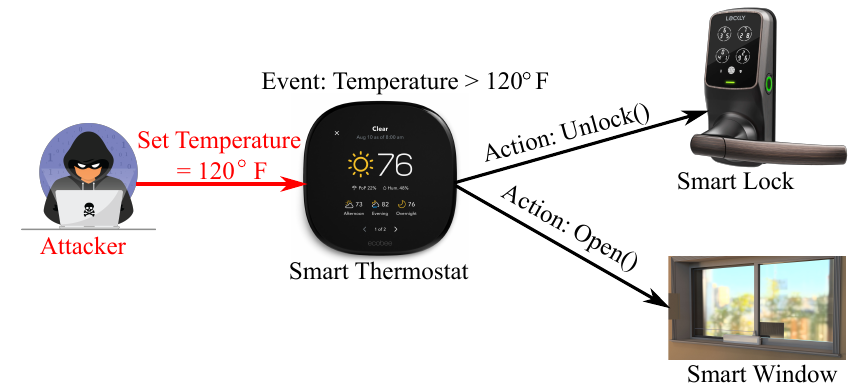}
    \vspace{-5pt}
    \caption{Remote injection attack scenario.}
    \label{fig:use_case_scenario}
    \vspace{-5pt}
\end{figure}

To implement a remote injection attack exploiting the chain of interactions, an attacker must inject fake event conditions into an IoT network and invoke relevant actions in target IoT devices. For example, suppose the owner of a smart home has set a customized rule:  "\textit{If the smart thermostat detects a temperature $\geq 120\degree$F, then unlock the smart lock on the front door and open the smart window}". In this scenario, the attacker could inject a fake event that reports the temperature as $120 \degree$F, thereby maliciously triggering the \emph{unlock()} and \emph{open()} commands for the smart lock and smart window. Fig. \ref{fig:use_case_scenario} represents the example scenario.

We assume that the attacker injects a set of event conditions $C = \{c_i\}, 1 \leq i \leq n$, where $n$ is the number of event conditions required to compromise a goal node. The attacker performs injection actions through a set of exploits $\xi = \{e_j\}, 1 \leq j \leq m$, where $m$ is the number of exploits, and $m \leq n$. Each exploit $e_j$ represents an interaction between an attacker and an IoT hub that reports an event condition $c_i$ at time $t$. The attacker adopts an attack strategy to select a sequence of optimal exploits $\xi$ to apply them to the IoT hub\cite{alam2024iotwarden}. 

\subsection{Attack Characterization}
\label{attack_characterization}

A remote injection attack can be modeled using a Directed Acyclic Graph (DAG), $\mathcal{G} = \{C, \xi\}$ demonstrating causal relationships between event conditions and corresponding actions \cite{ammann2022-monotonicity}. Hence, event conditions act as graph \emph{nodes} while actions(i.e., exploits) act as \emph{edges}. The root of the graph represents the starting point of the attack injection and one of the leaf nodes represents the goal node. Note that the goal node is not fixed and can be different from time-to-time. When a defender generates a DAG for the purpose of analyzing the security of an IoT network, the defender must enforce the \emph{monotonicity property} \cite{ammann2022-monotonicity} in the DAG to avoid the \emph{state explosion problem} \cite{sheyner2002-attack-graph-scalability-issue}. This monotonicity property ensures that the prior nodes in the DAG have no impact on the successor nodes. The enforcement of this property keeps the graph reasonably small to perform additional security analysis. We assume that the defender generates attack graphs using available security tools (e.g., Topological
Vulnerability Analysis (TVA) \cite{jajodia2006-attack-graph-generation}) and conducts security analysis on constructed graphs. 

\subsection{Threat Model}


The attacker performs active reconnaissance to discover attack vectors in the network and remotely injects fake event conditions using software called \emph{ghost devices}\cite{alam2022_iotmonitor}. These ghost devices are capable of simulating the behavior of real IoT devices. The attacker discovers device credentials from public forums and manufacturer's websites and utilizes these credentials to impersonate legitimate IoT devices. In addition, the attacker can capture network traffic using sniffers and extract real-time event traffic information by performing packet analysis using different IoT network utilities\cite{Trimananda2020PacketLevelSF}, \cite{zhang2018_homonit}. The attacker can profile defense actions and have the ability to perform \emph{opportunistic attacks} \cite{sbirnbach2019_peeves}. 

In an online defense system, the defender profiles ongoing attack actions and takes optimal defense actions to minimize the impact of attack actions in the network. We presume that the attacker cannot compromise the defense system and always takes minimal injection actions to evade detection by the defense system. Furthermore, we assume that the attacker is unable to compromise the sensors which are used to collect verifiable physical evidence and the IoT hub through which all IoT communications occur.

\section{RESTRAIN:\underline{Re}inforcement Learning-based \underline{S}ecure Framework for \underline{Tr}igger-\underline{A}ction \underline{I}oT E\underline{n}vironment}
\label{restrain-system-overview}

\begin{figure}[htbp]
    \centering
    \includegraphics[width=0.95\linewidth]{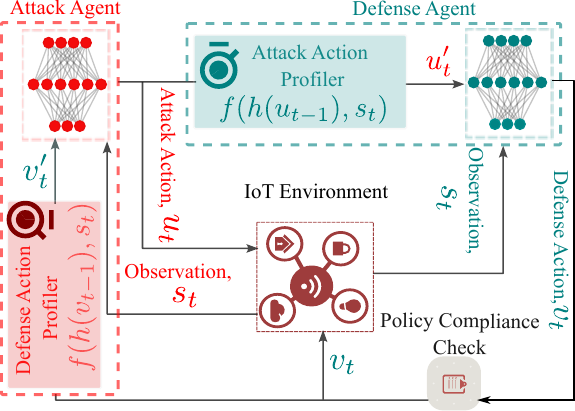}
    \caption{System overview of RESTRAIN. The attack agent and the defense agent take $u_t$ and $v_t$ actions, respectively, at current time-step based on the observation $s_t$ from the environment. The agents have the capability of profiling opponents' actions as their observation. The defense action goes through the \textit{policy compliance check} before applying to the IoT environment.}
    \label{fig:system_overview}
    \vspace{-5pt}
\end{figure}

We present RESTRAIN as a multi-agent RL framework where a defense agent maximizes its security gain by optimizing a defense policy and an attack agent learns strategies to take best injection actions. Fig. \ref{fig:system_overview} shows the system architecture of RESTRAIN. The attack agent observes the IoT environment to get current environment state as well as profiles the defense actions to estimate the latest defense action applied to the environment. The attack agent exploits these information to take current optimal action to the environment. Similarly, the defense agent observes the up-to-date security state of the environment and profiles the attack actions to determine the best defense policy. Unlike the attack agent, the defense agent recommends the defense action to the IoT hub rather than directly applying to the IoT environment. Then, the hub runs a \emph{policy compliance check} against predefined security policies for the defense action. Finally, the hub applies the defense action to the IoT environment if it passes the policy compliance check.


\subsection{System Environment}
We consider a smart home network as the system environment where the security states of the IoT devices, a set of attack actions, and a set of defense actions are encoded in a \textit{finite state machine}. The state machine consists of $\mathcal{N_S}$ unique states, $S = \{s_i\}, 1 \leq i \leq \mathcal{N_S}$, $\mathcal{N_U}$ unique attack actions, $U = \{u_j\}, 1 \leq j \leq \mathcal{N_U}$, and $\mathcal{N_V}$ unique defense actions, $V = \{v_k\}, 1 \leq k \leq \mathcal{N_V}$. Hence, $s_t$ represents the security state of the environment at time $t$ after both the attack agent and the defense agent perform the actions $u_{t-1}$ and $v_{t-1}$, respectively, at time $t-1$. We define $s_t$ as the security state of the environment once the event condition $c_t$ is reported to the hub. Note that the state $s_t$ represents the input for the RL algorithms used for optimizing the policies for the defense agent and the attack agent. The environment transitions from $s_{t-1}$ to $s_t$ with probability of $T(s_{t-1}, u_{t-1}, s_t) = Pr(s_t | s_{t-1}, u_{t-1})$ due to the attack action $u_{t-1}$. If the transition occurs due to the defense action $v_{t-1}$, it occurs with probability $T(s_{t-1}, v_{t-1}, s_t) = Pr(s_t | s_{t-1}, v_{t-1})$. For a successful transition at time $t$, the environment returns $r_{u_t}$ and $r_{v_t}$ as rewards to the attack agent and the defense agent, respectively. 

\subsection{Attack Agent}
The attack agent performs the following two types of actions (i.e., $\mathcal{N_U} = 2$) to implement a remote injection attack: 1) $u_r$: performing \textit{active reconnaissance} to discover attack vectors in the network; and 2) $u_i$: performing \textit{event injection} to report fake event conditions to the hub to invoke unauthorized actions in target IoT devices. The attack agent infers the latest defense action, $v_t'$ by conducting opponent modeling $v_t' = f \big( h(v_{t-1}), s_t \big)$ using an LSTM-based Recurrent Neural Network \cite{hochreiter1997long}, where $h(v_{t-1}) = \hspace{0.1cm} <v_1, v_2, ..., v_{t-1}>$. The objective is to optimize a policy $\pi^*_{u_t}$ to take an optimal sequence of actions (e.g., $u_r$, $u_i$) maximizing the overall discounted reward $\sum^T_{t=0} \gamma_u^t r_{u_t}$, where $\gamma_u \in [0,1)$ determines how much the future reward should be discounted from the current time step $t$.  

\smallskip
\subsubsection{Reward Function}
The objective of the attack reward function is to take optimal actions. We design the attack reward function in such a way that the attack agent takes minimal injection actions so that it does not expose itself to the defense agent by taking aggressive injections. In other words, the attack agent prefers more reconnaissance actions. At time $t$, the attack agent selects the action $u_t$ following the reward function as given in equation \eqref{eqn:reward_function-attacker}:
\begin{equation} 
    r_{u_t} = 
    \begin{cases}
        r_{u_r} - \lambda log(\kappa_u) & \mbox{ if } u_t = u_r, \\ 
        r_{u_i} + log(n_i) - \lambda n_i + r_{g} & \mbox{ otherwise. } 
    \end{cases}
    \label{eqn:reward_function-attacker}
\end{equation}

\noindent where $r_{u_t}$ represents the reward that the attack agent receives from the environment for taking the action $u_t$ at time $t$. $r_{u_r}$ represents the reward for taking the reconnaissance action, $u_r$. On the other hand, $r_{u_i}$ represents the reward for injecting a fake event condition using the action $u_i$ into the hub. $r_{g}$ is the reward the agent receives if it is successful in injecting an event condition that helps invoke an undesired action in the goal node. At any given time step, the parameter $\kappa_u$ denotes the number of consecutive reconnaissance actions, $u_r$ performed by the attack agent. The parameters $\lambda$ denotes \textit{attack proximity factor} that signals a defense agent how close the attack agent is to the goal node, and $n_i$ represents the number of injection actions performed thus far. The attack agent is incentivized to minimize both these parameters, as the larger values of these parameters make the attack actions increasingly more suspicious to the defense agent. We compute $\lambda \in [0, 1)$ as a ratio between the attack agent's current position and the goal node's position in the attack sequence. 

\subsection{Defense Agent}
The objective of the defense agent is to obtain an optimal \textit{defense policy} to recommend real-time defense actions to the smart hub maximizing the overall security gain. The defense agent performs two types of actions (i.e., $\mathcal{N_V} = 2$): 1) $v_a$: \textit{assessing} the security state of the environment, and 2) $v_b$: \textit{blocking} unsafe trigger conditions. The defense agent takes $v_a$ to assess how many event conditions $\{c_i\}$ the attack agent already injected into the hub. On the other hand, the defense agent performs action $v_b$ to set a lower bound of the injection actions required to compromise a node by the attack agent. The defense agent incorporates opponent's modeling in its decision process and infers the latest attack action. In order to infer the latest attack action, the defense agent utilizes a history of attack actions as follows:  $u_t' = f \big(h(u_{t-1}), s_t \big)$, where $h(u_{t-1}) = \hspace{0.1cm} <u_1, u_2, ..., u_{t-1}>$. Like the attack agent, the defense agent optimizes defense policy $\pi^*_{v_t}$ yielding a sequence of $v_a$ and $v_b$ maximizing $\sum^T_{t=0} \gamma_v^t r_{v_t}$, where $\gamma_v \in [0,1)$ is the discount factor.\\

\subsubsection{Reward Function}
The reward function is designed to motivate the defense agent to take optimal actions to provide RESTRAIN with active defense against the attack agent without hurting the availability of the network devices. To select an optimal action $v_t$ at time $t$, the defense agent utilizes the reward function as given in equation \eqref{eqn:reward_function-defender}:
\begin{equation} 
    r_{v_t} = 
    \begin{cases}
        r_{v_a} - \omega_d log(\sigma \kappa_v) + \lambda \sigma & \mbox{ if } v_t = v_a, \\ 
        r_{v_b} - \sigma - log(n_b) & \mbox{ otherwise}.
    \end{cases}
    \label{eqn:reward_function-defender}
\end{equation}

\noindent where $r_{v_t}$ is the reward the defense agent receives for taking the action $v_t$. $r_{v_a}$ and $r_{v_b}$ represents the reward the defense agent receives from the environment for the security assessment actions ($v_a$) and the reward given for the block actions ($v_b$), respectively. The parameter $\kappa_v$ denotes the number of consecutive $v_a$ actions performed by the defense agent, and the parameter $\sigma \in (0,1]$ denotes \textit{injection threshold} that quantifies the tolerance level of the defense agent against injection actions, $u_i$. We define $\sigma$ using the following formula: $\sigma = 1 - \lambda$. When $\sigma$ is closer to $1$, the defense agent becomes lenient against injection actions and prefers to take $v_a$ actions more often. Once $\sigma \leq \lambda$, the defense agent takes block actions $v_b$ aggressively to prevent the attack agent from reaching the goal node. The parameter $n_b$ in equation \eqref{eqn:reward_function-defender} represents the number of block actions taken by the defense agent thus far, which is required to minimize to maximize the reward $r_{v_t}$. The parameter $\omega_d$ is a weighting factor which is empirically chosen to control the impact of injection threshold in the reward function.

\section{Simulation Results}
\label{simulation-results}
\subsection{Experiment}
We develop an IoT environment using OpenAI Gym\cite{brockman2016openaigym} to implement RESTRAIN. In particular, a trigger-action IoT platform is developed complying with the OpenAI Gym structure to simulate real-time attack-defense strategies using reinforcement learning. An RL algorithm called Deep Recurrent Q-Network (DRQN\cite{hausknecht2015deep}) is implemented in Python leveraging the TensorFlow\cite{abadi2016tensorflow} framework. We train the RL models for both the defense agent and the attack agent using a machine equipped with Apple M3 Pro chip and 18GB RAM running on macOS Sonoma 14.4.


We utilize the PEEVES \cite{sbirnbach2019_peeves} dataset to extract event conditions to define the network structure of RESTRAIN. This dataset records IoT event transactions from 12 different event sources and physical evidence measured by 48 sensors. Event transactions and physical evidence share causal relationships, and therefore, we use the measured physical evidence to verify the occurrence of those event transactions. We translate the verifiable event conditions into network states and encode them into an IoT system environment. We encapsulate reward functions into the environment to provide feedback to agents for their actions.

\begin{table}[htbp]
    \centering
    \caption{optimal hyperparameters}
    \label{tab:hyperparameter}
    \begin{tabular}{|c|c|}
        \hline
        discount factor, $\gamma$ & $0.99$ \\
        \hline
        batch size & $32$ \\
        \hline
        learning rate, $\alpha$ & $0.001$
        \\
        \hline
        $(\epsilon, \epsilon_{decay}, \epsilon_{min})$ & $(1.0, 9995,0.005)$\\ \hline
        target network update frequency, $C$ & $10$ episodes\\ \hline
        replay buffer size & $5k$ \\ \hline
        activation function for Dense layers & \textit{ReLU} \\ 
        \hline 
        activation function for LSTM layers & \textit{Tanh} \\ 
        \hline
        weighting factor, $\omega_d$ & 0.01\\
        \hline
    \end{tabular}
    \vspace{-10pt}
\end{table}

We employ a \textit{function approximator} for each agent and conduct hyperparameter tuning to figure out an optimal set of hyperparameters that maximizes the agent's objective. Especially, we perform a grid search \cite{bergstra2011algorithms}, \cite{RLPG_Lokesh_2023} over a range of hyperparameters including \textit{learning rate}, \textit{batch size}, \textit{exploration rate $\epsilon$}, and \textit{network complexity}, etc. Table \ref{tab:hyperparameter} enlists the optimal set of hyperparameters obtained through the grid search approach that we use to train our agents. As we model our IoT trigger action platform as a multi-agent learning system, we use the same set of parameters to facilitate the fair competition between the attack agent and the defense agent. The neural network consists of four hidden layers. The first hidden layer is Dense layer with 64 neurons followed by two LSTM layers. Each LSTM layer has 32 neurons. Finally, the output of the final LSTM layer is passed through a Dense layer of 16 neurons before passing through the output layer. We train our \textit{function approximators} using a stochastic gradient descent based optimization algorithm named Adam\cite{kingma2014adam} optimizer and Mean Squared Error (MSE) as a loss function.

In our simulation, agents utilize the $\epsilon$\textit{-greedy} policy method defined in equation \eqref{eqn:epsilon_greedy} to accelerate the learning process\cite{gimelfarb2020epsilon}, \cite{das2021d}. 
\begin{equation} 
    \pi^\epsilon (a|s) = 
    \begin{cases}
        1-\epsilon_t + \frac{\epsilon_t}{|A|} & \mbox{ if } a = \mbox{argmax}_{a'\in A}Q_t(s, a') \\ 
        \frac{\epsilon_t}{|A|}& \mbox{ otherwise}  \\ 
    \end{cases}
    \label{eqn:epsilon_greedy}
\end{equation}
where $a\in \{u, v\}$ represents actions for both the attack agent and the defense agent. In $\epsilon$\textit{-greedy} policy, $\pi^\epsilon$ chooses a random action from the action space $A$ with the probability of $\epsilon_t \in [0,1]$; otherwise, it exploits a greedy action according to the $Q_t$. 

\subsection{Performance Evaluation}
We run the simulation for $350$ episodes where each episode performs $50$ gradient updates. We evaluate RESTRAIN in terms of total reward, number of injection actions taken by the attack agent, and the number of block actions employed by the defense agent to safeguard the IoT network. We also consider \emph{attack injection threshold} and \emph{attack proximity factor} for modeling the attack-defense dynamic of RESTRAIN. Finally, we assess the computational overhead of our proposed approach in terms of time incurred on the network environment.

\subsubsection{Attack-Defense Reward}
\begin{figure}[htbp]
    \centering
    \subfloat[Attack agent reward.\label{1a}]{%
       \includegraphics[width=0.49\linewidth]{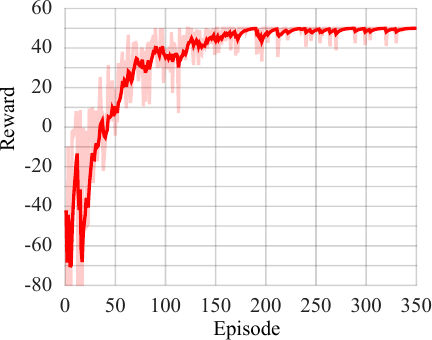}}
    \hfill
  \subfloat[Defense agent reward.\label{1b}]{%
        \includegraphics[width=0.49\linewidth]{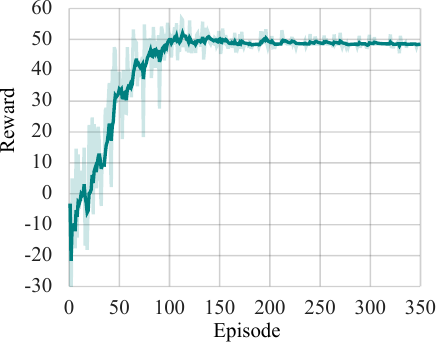}}
    \caption{The reward graphs for the attack agent and the defense agent.}
    \label{reward_graph}
    \vspace{-10pt}
\end{figure}

The reward of the attack agent represents the accumulated attack gain received by the attack agent from the environment for taking injection actions in all time steps in a certain episode. The environment provides a discrete reward to the attack agent computed using equation \eqref{eqn:reward_function-attacker} for each attack action. The attack agent tries to discern an optimal attack policy to maximize the accumulated reward that indicates the agent's capability to perform optimal injection actions. Fig. \ref{reward_graph}(a) shows the learning trend of the attack agent. At early stage of learning, the reward exhibits unstable pattern which means that the agent does not learn an optimal policy. As training progresses, the agent learns to take optimal actions. More specifically, the reward graph shows more stable pattern after $\approx 150$ episodes converging to the maximized cumulative reward.

Similarly, the defense agent achieves security gain as a form of reward according to equation \eqref{eqn:reward_function-defender} from the environment at every time step for each action it takes and tries to maximize the accumulated reward to determine the optimal defense policy in an episode. The defense agent also exhibits the unstable pattern of the accumulated reward like the attack agent at the initial stage of learning. However, the defense agent figures out the optimal defense policy more faster than the attack agent. Particularly, the defense agent achieves stable reward pattern at $\approx 100$ episodes and shows a convergence trend afterwards as shown in Fig. \ref{reward_graph}(b).\\


\subsubsection{Attack-Defense Dynamic}

\begin{figure}[htbp]
	\begin{center}
		\includegraphics[width=0.7\linewidth]{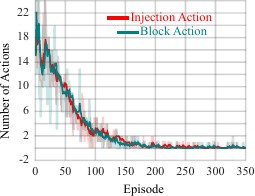}
        \vspace{-5pt}
		\caption{Injection-Block actions taken by attack and defense agents.\label{fig:injection_block_operations}}
        \vspace{-10pt}
	\end{center}
\end{figure}

As we discussed in Section III, the attack agent and the defense agent make changes in the environment states by taking \textit{injection} and \textit{block} actions, respectively. To model the attack-defense dynamic, we take into account how an agent reacts to the opponent agent's action for which a state transition occurs. Fig. \ref{fig:injection_block_operations} depicts the attack-defense dynamic.

When the attack agent exploits the network by taking more injection actions, the defense agent tries to secure the network by taking more block actions. However, as the more injection actions increase the chances of exposing the attack agent to the defense agent and the more block actions constraints the network availability, both agents try to take optimal injection and block actions as required in order to reach the goal node or protect the goal node. From Fig. \ref{fig:injection_block_operations}, it can be observed that at the beginning of the training, when the attack agent exploits the network with more injection actions, the defense agent also takes more block actions to compete against the attack agent's aggressiveness. In contrast, the agents prefer taking \emph{reconnaissance} and \emph{assessing} actions more frequently to \emph{injection} and \emph{block} actions.


\subsubsection{Defense Tolerance Level vs Attack Proximity Factor}

\begin{figure}[htbp]
    \centering
    \vspace{-5pt}
    \includegraphics[width=0.95\linewidth]{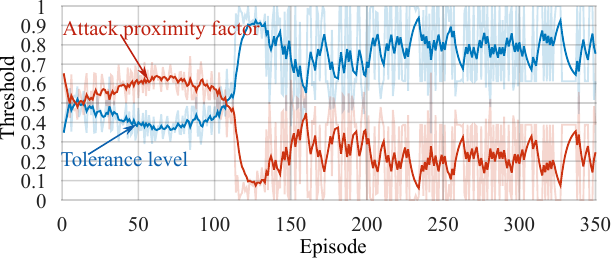}
    \vspace{-5pt}
    \caption{The defense tolerance level vs the attack proximity factor.}
    \label{fig:injection_proximity}
    \vspace{-5pt}
\end{figure}
We define the \emph{attack proximity factor} as a metric to determine how close the attack agent is to the goal node whereas the \emph{tolerance level} to represent the injection threshold that tells a defense agent when to become aggressive against attack activities. In RESTRAIN, the increase in attack proximity factor leads to the decrease of tolerance level and vice versa. When the attack agent starts performing injection actions aggressively, the defense agent starts reducing its tolerance level so that it can take more aggressive defense actions to obstruct the attack progression. Fig. \ref{fig:injection_proximity} displays the result. It is clearly evident from Fig. \ref{fig:injection_proximity} that the attack proximity factor increases at the initial stages of learning. This is because the defense agent allows the attack agent to take more injection actions so that it can learn the attack dynamic and gradually reduce its tolerance level. The defense agent starts taking aggressive defense actions (i.e., \emph{block} actions) when the tolerance level becomes smaller or equal to the attack proximity factor. Once the defense agent blocks the crucial triggers from the IoT environment, it becomes infeasible for the attack agent to progress towards the goal node even though the tolerance level is increased a lot. This phenomenon is evident in Fig. \ref{fig:injection_proximity}. When the tolerance level increases after $\approx 110$ episodes, the attack agent cannot increase the attack proximity factor beyond $0.5$ demonstrating the fact that it becomes harder for the attack agent to inject malicious triggers into the environment to invoke invalid actions to reach to the goal node. 

\subsubsection{Computational Overhead}
\begin{figure}[htbp]
	\begin{center}
        \vspace{-5pt}
		\includegraphics[width=0.7\linewidth]{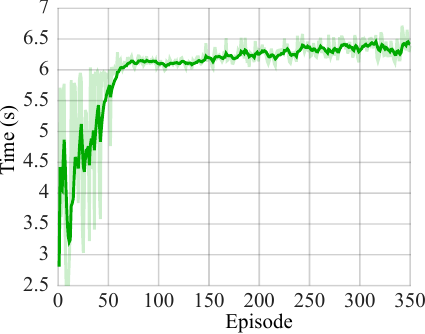}
        \vspace{-5pt}
		\caption{Computational Overhead. \label{fig:computational_overhead}}
	\end{center}
        \vspace{-10pt}
\end{figure}

The enforcement of security mechanism always incurs computational burden on systems and networks. RESTRAIN also adds some computational overhead on the IoT environment. To quantify this overhead, we compute the time (in seconds) the reinforcement learning algorithm takes in each episode to train the agents. Fig. \ref{fig:computational_overhead} represents the computational overhead of RESTRAIN over episodes. It is evident that RESTRAIN learns to stabilize the computational overhead (e.g., $\leq 6.5s$) after a certain number of episodes (e.g., after $\approx 75$ episodes). This stability in computational overhead signifies the applicability of the proposed system in real IoT settings.

\section{Conclusion}
\label{conclusion}
In this paper, we propose an RL-based multi-agent real-time defense system called RESTRAIN for trigger-action IoT environment which facilitates a defense agent to optimize its defense policy by modeling attack strategies and an attack agent to optimally inject fake event conditions to invoke invalid actions in target IoT devices. We design novel reward functions and train both agents using Deep Recurrent Q-Network (DRQN) by creating a custom IoT environment using OpenAI Gym framework. We run extensive simulations to evaluate our proposed framework. Simulation results exhibit that the defense agent can effectively safeguard IoT environment from complex remote injection attacks with minimal computational overhead. 

Deploying RESTRAIN in cloud and  evaluating our proposed solution using real-world testbed is our future work. Additionally, we will integrate user-configurable overhead in the defense system to enable users to incorporate contextual computational constraints.

 \section*{Acknowledgment}
This work is supported in part by DoD Center of Excellence in AI and Machine Learning (CoE-AIML) under Contract Number W911NF-20-2-0277 with the U.S. Army Research Laboratory, National Science Foundation under Grant No. 2219742 and Grant No. 2131001.

\bibliographystyle{./References/IEEEtran}
\bibliography{./References/icc25-bibliography}

\end{document}